\documentclass[aoas,MSNbibl,nameyear,dvips]{arximspdf}
\usepackage{graphicx}
\doi{10.1214/13-AOAS651} 
\volume{7}
\issue{3}
\pubyear{2013}
\firstpage{1286}
\lastpage{1310}

\makeatletter

\newproclaim{definition}[theorem]{Definition}
\newproclaim{example}[theorem]{Example}

\newcommand{\M}{\mathrm{M}}
\newcommand{\Cov}{\operatorname{Cov}}
\newcommand{\real}{\mathbb{R}}
\newcommand{\myE}{\mathbb{E}}
\newcommand{\bs}{\mathbf{s}}
\newcommand{\bw}{\mathbf{w}}
\newcommand{\bY}{\mathbf{Y}}
\newcommand{\bU}{\mathbf{U}}
\newcommand{\bV}{\mathbf{V}}
\newcommand{\bX}{\mathbf{X}}
\newcommand{\bD}{\mathbf{D}}
\newcommand{\bL}{\mathbf{L}}
\newcommand{\bH}{\mathbf{H}}
\newcommand{\bZ}{\mathbf{Z}}
\makeatother

\begin{document}
\begin{frontmatter}

\title{Parameter tuning for a multi-fidelity dynamical model of the
magnetosphere\thanksref{TT1}}
\runtitle{Parameter tuning dynamical computer models}
\thankstext{TT1}{Supported by NSF Grant AGS-0934488.}

\begin{aug}
\author[A]{\fnms{William} \snm{Kleiber}\corref{}\ead[label=e1]{william.kleiber@colorado.edu}\thanksref{t1}},
\author[B]{\fnms{Stephan R.} \snm{Sain}\thanksref{t2}}, %
\author[B]{\fnms{Matthew J.} \snm{Heaton}\thanksref{t2}}, %
\author[C]{\fnms{Michael} \snm{Wiltberger}\thanksref{t2}}, %
\author[D]{\fnms{C.~Shane} \snm{Reese}\thanksref{t3}} %
\and
\author[E]{\fnms{Derek} \snm{Bingham}\thanksref{t4}} %
\runauthor{W. Kleiber et al.}
\affiliation{University of Colorado\thanksmark{t1}, Brigham Young University\thanksmark{t2},\break Brigham Young
University\thanksmark{t3} and Simon Fraser University\thanksmark{t4}}
\address[A]{W. Kleiber\\
Department of Applied Mathematics\\
University of Colorado\\
Boulder, Colorado\\
USA\\
\printead{e1}} 
\address[B]{S. R. Sain\\
M. J. Heaton\\
Institute for Mathematics Applied\\
\quad to Geosciences\\
National Center for\\
\quad Atmospheric Research\\
Boulder, Colorado\\
USA\vspace*{24pt}}
\address[C]{M. Wiltberger\\
High Altitude Observatory\\
National Center for\\
\quad Atmospheric Research\\
Boulder, Colorado\\
USA}
\address[D]{C. S. Reese\\
Department of Statistics\\
Brigham Young University\hspace*{30pt}\\
Provo, Utah\\
USA}
\address[E]{D. Bingham\\
Department of Statistics and\\
\quad Actuarial Science\\
Simon Fraser University\\
Burnaby, BC\\
Canada}
\end{aug}

\received{\smonth{12} \syear{2012}}

%
\begin{abstract}
Geomagnetic storms play a critical role in space weather physics with
the potential for far reaching
economic impacts including power grid outages, air traffic rerouting,
satellite damage
and GPS disruption. The LFM--MIX is a state-of-the-art coupled
magnetospheric--ionospheric
model capable of simulating geomagnetic storms. Imbedded in this model are
physical equations for turning the magnetohydrodynamic state parameters
into energy and flux of electrons entering the ionosphere, involving a
set of input
parameters. The exact
values of these input parameters in the model are unknown, and we seek
to quantify the
uncertainty about these parameters when model output is compared to
observations.
The model is available at different fidelities: a lower fidelity which
is faster to
run, and a higher fidelity but more computationally intense version.
Model output and observational data are large spatiotemporal systems;
the traditional design and analysis of computer experiments is unable to
cope with such large data sets that involve multiple fidelities of
model output.
We develop an approach to this inverse problem for large
spatiotemporal data sets
that incorporates two different versions of the physical model. After
an initial design,
we propose a sequential design based on expected improvement. For the LFM--MIX,
the additional run suggested by expected improvement diminishes posterior
uncertainty by ruling out a posterior mode and shrinking the width of the
posterior distribution.
We also illustrate our approach using the Lorenz `96 system of equations
for a simplified atmosphere, using known input parameters. For
the Lorenz `96 system, after
performing sequential runs based on expected improvement, the posterior
mode converges to the true value and the posterior variability is reduced.
\end{abstract}

%

\end{frontmatter}

\section{Introduction} \label{secintroduction}

The Lyon--Fedder--Mobarry (LFM)
magnetohydrodynamical model, coupled with the MIX model for the ionosphere,
creating the coupled LFM--MIX, is a state-of-the-art physical model for
geomagnetic storms occurring in near-Earth space [\citet{lyon2004}].
The LFM--MIX is used to explore and understand the physics of
space weather, and is a crucial part of an ongoing effort to
build a space weather forecasting system.
The LFM--MIX contains three input parameters embedded in physical
equations for
turning the LFM state parameters into energy and flux [\citet{wiltberger2009}].
Exact values of these input
parameters are unknown, and our goal is to quantify the uncertainty surrounding
these parameters when model output is compared to an observed storm,
posing substantial statistical challenges including large spatiotemporal
systems of observations and model output, as well as the need to
incorporate multiple versions of the LFM--MIX.

\subsection{Geomagnetic storms} \label{secintrostorms}

Geomagnetic storms play an increasingly important role in society.
A recent National Academy of Sciences report outlined past
occurrences of geomagnetic storm disruptions, and discussed the
importance of
preparedness in the future when the Sun returns to its solar peak in 2013,
which leads to larger and more frequent geomagnetic storms
[\citet{nas2008}].
Intense geomagnetic storms adversely affect satellites and can have significant
associated costs; in 1994 a Canadian telecommunication satellite experienced
an outage due to a strong storm, and recovery of the satellite cost
between \$50 million and \$70 million. Large storms can interact with
electric grids;
a superstorm in March 1989 shut off electricity to the province of Qu\'
ebec, Canada
for nine hours. Global position systems (GPS) and communication systems are
affected by large storms; the Federal Aviation Administration's Wide Area
Augmentation System (WAAS) is a GPS location system for aircraft, whose
vertical navigation system was shut down for approximately 30 hours in 2003
due to a series of powerful storms. As society has become increasingly reliant
on electricity and satellite communication, the potential devastating effects
of geomagnetic storms are magnified.

Geomagnetic storms are caused by the interaction of the plasma and magnetic
field of the Sun interacting with Earth's magnetic field. Coronal Mass
Ejections (CMEs) from the Sun release massive twisted magnetic field
configurations that can deposit substantial energy in the region of
near-Earth space known as the magnetosphere. The energy is stored
for a while, and then is released in an explosive fashion,
sending particles down magnetic field lines into the ionosphere
causing the aurora borealis or northern lights.

\subsection{Computer experiments} \label{secintrodace}

In the computer experiments literature, the tuning of physical model
parameters to observations is called an inverse problem,
and is sometimes referred to as a calibration problem
[\citet{santner2003,tarantola2005}].
Two features of our setup make the traditional approach
to design and analysis of computer experiments infeasible.
First, observational data and computer model output are highly
multivariate; modeling model output and observations as realizations
from a Gaussian process [e.g., as popularized by \citet{sacks1989},
see also \citet{kennedy2001} and \citet{higdon2004}] is impractical due
to the dimensionality of the covariance matrix.
The second issue is that the LFM--MIX is available at multiple fidelities.
In particular, solving the physical equations making up the LFM at a lower
resolution yields model output that is jointly faster to calculate but does
not match up as well with observations, a version we call low fidelity.
Alternatively, at a higher resolution the LFM yields output whose
spatial features are more consistent with observational data,
but which takes
substantially longer to run (approximately an eightfold increase in
computation time), a version we call high fidelity. We aim
to exploit a statistical link between the model fidelities, thereby
allowing us to explore the input parameter space using the cheaper low fidelity
version, while performing fewer runs of the high fidelity version.

The problem of high-dimensional observations and model output has
recently become acknowledged in the computer experiments literature.
\citet{higdon2008} recommend decomposing model output and model bias terms
as weighted sums of orthogonal basis functions. The weights on the basis
functions are then modeled as Gaussian processes. Indeed, the notion of
an orthogonal decomposition has been further used by various authors to
reduce the high dimensionality of vector-valued model output [\citet
{higdon2008cmame,wilkinson2010}]. \citet{pratola2013} introduce a fast
approach to
calibration for large complex computer models.
In the geophysical sciences, model output is
often spatiotemporal in nature, which typically gives rise to large
data sets. \citet{bhat2010} develop a calibration
approach for multivariate spatial data, modeling the model output as
a Gaussian process across space and input setting, exploiting a
separable covariance structure. Our model and data also evolve across time,
and the presence of multiple fidelities of model output challenge the
approach of \citet{bhat2010}.

Accounting for multiple versions of model output is a second problem
that has recently arisen in the computer experiments literature.
\citet{kennedy2000} introduce an autoregressive Markov property for multiple
fidelities of model output, modeling the innovation as a Gaussian process.
While their idea is extended to a continuum of model fidelities, a crucial
and restrictive assumption is that the model output is scalar.
\citet{qian2006} develop an approach to combining two levels of fidelity
that is extended to a Bayesian hierarchical setting by \citet{qian2008}.
The idea is to decompose the high fidelity
output as a regression on the low fidelity version, and model the intercept
and slope as Gaussian processes. \citet{forrester2007} and \citet
{legratiet2012}
recommend co-kriging for multiple fidelities of output, but do not
consider the issue of large data sets. We exploit similar ideas to these
authors in our construction, although we must take care to reduce the
dimensionality of the data, as both versions of the LFM--MIX are highly
multivariate. It is worth mentioning that there is some literature on
emulators for multivariate computer models, but our current interest
is not in emulation, but rather parameter identification [\citet
{rougier2008,rougier2009}].

Herein we develop methodology for quantifying the uncertainty about tuning
parameters for high-dimensional spatiotemporal
observations and the physical model with two levels of fidelity. We
exploit an empirical orthogonal function (EOF) decomposition of the low
fidelity spatial field, and an EOF decomposition of a discrepancy function
linking the low and high fidelity versions of the computer model.
Our work generalizes that of \citet{kennedy2001} to account for large
spatiotemporal data sets. The techniques introduced below also
generalize the
approach of
\citet{higdon2008} to account for two levels of model fidelity. The
methodology is illustrated on the LFM--MIX and the Lorenz `96 system of
equations governing a simplified atmosphere [\citeauthor{lorenz1996} (\citeyear{lorenz1996,lorenz2005})],
where we know true values of the input parameters. For both models,
after initial parameter estimation, we propose a sequential design
based on
expected improvement [EI, \citet{jones1998}]. Our development of expected
improvement generalizes the approach of \citet{jones1998} to sequential
design for
spatiotemporal data.

\section{LFM--MIX and observations} \label{seccomputerobservations}

The physical model we examine is a coupled magnetospheric--ionospheric model
for geomagnetic storms in near-Earth space. The magnetohydrodynamical
solver is the Lyon--Fedder--Mobarry (LFM) model which consists of five physical
equations defining the spatial and temporal evolution of the interaction
between the solar wind and Earth's magnetosphere. These five magnetohydrodynamic
equations must be solved numerically by discretizing the equations to a
spatiotemporal grid, using the partial donor method [\citet{wiltberger2004}].
There is a coarsest grid on which the equations are
solved that still yields physically meaningful model output at a reduced
computational cost. Discretizing the equations on a finer grid by doubling
the number of spatiotemporal points (in the polar and azimuthal angle
directions, as well as at a finer temporal scale) results in higher fidelity
model output, but substantially increases the computational time required
to complete model runs. Intuitively,
doubling the grid density in three directions results in a $2^3 = 8$-fold
increase in computation time; in practice, the higher resolution version
is an approximately $5.5$ to $6$-fold increase in computation time as compared
to the lower resolution. As
boundary conditions, the LFM requires solar wind, initial strength of the
magnetic field, and the level of ultraviolet light from the Sun. For
any single geomagnetic storm, these boundary conditions are fixed and
are not considered
input parameters.

The LFM solver is coupled to an ionospheric model, the MIX, forming
the fully coupled LFM--MIX. The MIX model requires information about the
energy and number flux of the electrons precipitating into the ionosphere
along magnetic field lines. Three physical equations define
energy and number flux inputs. The
equations relate initial energy $\varepsilon_0$, sound speed $c_s^2$,
number flux $F_0$, the density of innermost cells of the magnetospheric
grid $\rho$, the field aligned electrical potential energy difference
$\varepsilon_\|$, and upward field aligned current $J_\|$ as
%
%
\begin{equation}
\label{eqabr} \varepsilon_0 = \alpha c_s^2,\qquad
F_0 = \beta\rho\sqrt{\varepsilon_0}, \qquad\varepsilon_\| =
\frac{R J_\| \sqrt{\varepsilon_0}}{\rho};
\end{equation}
see \citet{wiltberger2009} for further discussion.
An important quantity called total energy is defined as $\varepsilon_0 +
\varepsilon_\|$.
Here, $\alpha, \beta$, and $R$ are tuning factors that are included
to account
for physical processes outside the scope of the LFM.
The exact values of these parameters are unknown,
and we seek to quantify the uncertainty about these parameters when model
output is compared to observations. The parameter $\alpha$ accounts for
the effects of calculating
electron temperature from the single fluid temperature, $\beta$ is included
to adjust for possible plasma anisotropy and controls a loss filling cone,
while $R$ allows scaling of the parallel potential drop based on the sign
of the current and accounts for the possibility of being outside the
regime of
the scaling. Notice the total energy is a nonlinear function of $\alpha$
and~$R$, while flux is a function of $\beta$; later when we
develop the statistical model, we take advantage of these functional
relationships.

Regardless of the resolution of the LFM input, the MIX coupler output
is always on
the same spatiotemporal resolution. Hence, unlike uncoupled models, the
low and high resolution
LFM--MIX output is co-located, and we will refer to the low resolution output
as low fidelity, and the high resolution output as high fidelity.
This allows us to directly compute the scalar difference between the
two fidelities without regridding.
Model output from the LFM--MIX is a bivariate spatiotemporal
field, for the variables of energy (in keV) and flux
(in $\frac{1}{\mathrm{cm}^2\mathrm{s}}$). Developing a bivariate spatiotemporal
model is beyond the scope of the current manuscript, and we focus on
uncertainty estimation using only the energy model output.

The observational data set we examine is a bivariate spatiotemporal field
observed during a January 10, 1997, geomagnetic storm from 2~pm to 4~pm
UTC, with 18
equally spaced time points. The storm was observed by the Ultraviolet
Imager on
the Polar satellite, deriving the two variables of energy (in keV) and
energy times flux
(in $\frac{\mathrm{mW}}{\mathrm{m}^2}$) simultaneously. The observations
were recorded on a grid of 170 locations, leading to a data set of 6120
correlated
observations. The LFM--MIX model output is on a grid of 1656 locations such
that the observational grid is a subset of the model output.

\section{Parameter estimation for the LFM--MIX} \label{seccalibration}

We require initial runs of the low and high fidelity model to inform a
statistical
relationship between the two. As our initial experimental design, we
run the LFM--MIX at a
sampling of points in the three-dimensional space defined by $\alpha
\in[0,0.5],
\beta\in[0,2.5]$, and $R \in[0,0.1]$, which is the hyperrectangle
defining physically feasible values of $(\alpha,\beta,R)$.

\subsection{Design}

Using the hyperrectangle $[0,0.5] \times[0,2.5] \times[0,0.1]$
of values for $\theta= (\alpha,\beta,R)$,
we ran the low fidelity version at 20 sets of input settings based on
a space-filling design [\citet{johnson1990}].
Call this model output $L(\bs,t,\theta_p)$ at
location $\bs\in\real^2$,
time $t$, and input setting $\theta_p = (\alpha_p,\beta_p,R_p),
p=1,\ldots,20$. We also ran the high fidelity version at a nested,
space-filled subset of 5 of the original 20. Similar to the low
fidelity, call
the model output $H(\bs,t,\theta_p)$, for $p=1,\ldots,5$.
Setting up the initial design in such a way that the low and high fidelity
versions are nested, that is, run at co-located input parameter settings,
yields direct observations of the discrepancy $H(\bs,t,\theta_p) -
L(\bs
,t,\theta_p)$,
and assists in developing the statistical relationship between the two. If
the design were not nested, we would require estimated discrepancies
$H(\bs,t,\theta_p) - \hat{L}(\bs,t,\theta_p)$ or
$\hat{H}(\bs,t,\theta_p) - L(\bs,t,\theta_p)$ to explore the statistical
relationship, thereby introducing additional uncertainty. The choice of 20
and 5 runs for the low and high fidelity model, respectively, is due to
the expensive computational cost of running the LFM--MIX. For our study
geomagnetic storm, the low fidelity model runs in 16 hours, while the
high fidelity
model requires approximately 84 hours per run on a Linux cluster with 8
processors.
In total, the initial
design took approximately 740 hours to run. Note the benefit of exploiting
the lower fidelity, but faster running version---had we run the high
fidelity model on the initial design of 20 input settings, the computational
time would be approximately 1680 hours. Hence, the inclusion of the
cheaper low fidelity model allows us to reduce the initial
computational load by
about $56\%$.

\subsection{Statistical model}

Following an approach popularized by \citet{kennedy2001}, we suppose
there is an
unknown setting, $\theta_0$, for which the high fidelity model is an
adequate representation of reality. In particular, for observations
of energy (in keV), $Y(\bs,t)$, at grid point $\bs$ and time $t$, we have
%
%
\begin{equation}
\label{eqobs} Y(\bs,t) = H(\bs,t,\theta_0) + \varepsilon(\bs,t),
\end{equation}
where $\varepsilon(\bs,t)$ is measurement error, which we assume to
be normally
distributed with mean zero and variance $\tau^2$. Our approach slightly
differs from \citet{kennedy2001} in that we do not entertain a model
discrepancy term.
Our setup is a large-scale inverse problem, where model discrepancy is not
part of the traditional setup [\citet{tarantola2005}]. We also point
out that
we have only one geomagnetic storm, and any model bias term would be
confounded with the error process $\varepsilon(\bs,t)$, without severe
simplifying assumptions.

To fully exploit the information from the low fidelity model, we require
a link between the coarse model $L$ and the higher fidelity model $H$, which
yields output fields that are more consistent with observational data.
Specifically, we link the low and high fidelity models with an additive
discrepancy function
$\delta(\bs,t,\theta)$, where
%
%
\begin{equation}
\label{eqreslink} H(\bs,t,\theta) = L(\bs,t,\theta) + \delta(\bs
,t,\theta).
\end{equation}
\citet{qian2008} considered including a multiplicative discrepancy function
as well, yielding a decomposition of the form $H(\bs,t,\theta) =
\gamma
(\bs,t,\theta) L(\bs,t,\theta) +
\delta(\bs,t,\theta)$. For the LFM--MIX, both fidelities produce
output fields that are of approximately the same magnitude, so we
consider only
an additive discrepancy function, although the greater flexibility of a full
multiplicative and additive bias may be useful in other settings. By
defining a statistical relationship between the low and high fidelity versions
of the LFM--MIX, we have inherently also developed an emulator for the
high fidelity model, based on runs from the cheaper low fidelity version,
but reassert that our main interest is in the parameters $(\alpha
,\beta,R)$.

The model and observations are highly multivariate space--time fields,
where, with only one storm and 20${}+{}$5 initial computer model runs, we have
748,260 correlated points (1656 grid locations for the 25 LFM--MIX output
runs at 18 time points plus 170 observation locations over 18 time points).
The traditional
approach used by \citet{kennedy2001} is challenging to implement for
large space--time data sets, as this would require inverting a
covariance matrix of dimension
748,260$\times$748,260. Indeed, in their implementation, the
covariance matrix
would have to be inverted at each step of an MCMC procedure.
Hence, with spirit similar to \citet{higdon2008},
we use a principal component decomposition approach to
reduce dimensionality. In particular, we decompose
the low resolution model output and discrepancy function as weighted sums
of orthogonal spatial basis functions. In the geophysical
sciences, these spatial functions are known as empirical orthogonal
functions [EOFs; \citet{wikle2010}]. In particular, define the
spatial vectors $\bX(t_i,\theta_p) = (L(\bs_1,t_i,\theta_p),\ldots,
L(\bs_{n_s},t_{i},\theta_{p}))'$, where $n_s=1656$ is the total number
of grid points of
model output, $n_t=18$ is the number of time points,
$i=1,\ldots,n_t$ and $p=1,\ldots,20$. Define the $n_s \times(20
\times n_t)$
dimensional matrix
\[
\bX= \bigl[\bX(t_1,\theta_1),\bX(t_2,
\theta_1),\ldots,\bX(t_{n_t},\theta_{20})\bigr]
\]
so that each column is a spatial vector at a given time point and
input setting. The EOFs are the columns
of $\bU$, where we use the singular value decomposition
$\bX= \bU\bD\bV'$, and the EOF coefficients are contained
in $\bD\bV'$. In particular, there are $20 \times n_t$ EOFs,
each of which is length $n_s$. We perform a similar decomposition
for the discrepancy process $\delta(\bs,t,\theta) = H(\bs,t,\theta
) -
L(\bs,t,\theta)$, where there are $5 \times n_t$ EOFs, each of which is
length $n_s$.
Our motivation for decomposing the model output as basis functions over
space, rather than space--time, is driven by exploratory analysis.
In particular, the first main spatial mode of variation
of the low fidelity model output (i.e., the first EOF) exhibits a
magnitude with
a structured form that is similar to the physical equation (\ref{eqabr})
and whose magnitude modulates
up and down as the CME passes over the Earth. This aligns with
expert understanding of geomagnetic storms, as the effect of the CME passing
over the Earth is a period of increasing energy and flux, followed by a
decay to pre-storm conditions.

We statistically model the low fidelity model output as a truncated sum of
weighted EOFs,
%
%
\begin{equation}
\label{eqLEOF} L(\bs,t,\theta) = \sum_{e=1}^{n_L}
u_{Le}(\bs) v_{e}(t,\theta) + \varepsilon_{L}(
\bs,t,\theta)
\end{equation}
and similarly the discrepancy function as
%
%
\begin{equation}
\label{eqdeltaEOF} \delta(\bs,t,\theta) = \sum_{e=1}^{n_\delta}
u_{\delta e}(\bs) w_{e}(t,\theta) + \varepsilon_{\delta}(
\bs,t,\theta),
\end{equation}
where the $u$ basis functions are the EOFs contained in the $\bU$
matrices above, and the $v$ and $w$ coefficients are the loadings
contained in the $\bD\bV'$ matrices.
We choose sum limits of $n_L=3$ and $n_\delta=4$ to capture $99\%$ of
variability
of low fidelity model output, and $90\%$ of variability of the
discrepancy process,
respectively. To capture $99\%$ of variability for the discrepancy process,
for example, we would require the first 26 EOFs, which would detract
from a parsimonious formulation; \citet{higdon2008} also suggest that
a Gaussian process representation of high order basis function coefficients
tends to perform poorly in terms of prediction.
Here, $\varepsilon_{L}$ and $\varepsilon_{\delta}$ are independent
mean zero
normally distributed white noise
error terms with variances $\tau_{L}^2$ and $\tau_{\delta}^2$,
respectively. The statistical model is completed by assuming the
coefficient processes
$v_{e}(t,\theta)$ and $w_{e}(t,\theta)$ are Gaussian processes.

Based on the physical equations that define the total energy and number
flux of precipitating electrons for the MIX model, we impose a
nontrivial mean
function on the first low fidelity loading, $v_{1}$. Utilizing the
functional form of the
total energy equation, $\varepsilon_0 + \varepsilon_\|$, we specify a nonlinear
mean function
%
%
\begin{equation}
\label{eqmeanfunction} \myE v_{1}(t,\theta) = \gamma_{0}
+ \gamma_{1} \alpha+ \gamma_{2} R \sqrt{\alpha} +
\gamma_{3} \cos(2\pi t/n_t) + \gamma_{4} \sin(2
\pi t/n_t).
\end{equation}
The harmonics in the mean function are due to the nature of geomagnetic storms;
as the CME passes over the Earth, the average background energy field
increases in magnitude followed by a decay to the average background.
The harmonics capture the physical temporal evolution of the geomagnetic
storm over the period of our observations. We give the $w_1$ loading process
a constant mean parameter, allowing the variability of the discrepancy process
across input setting to be captured by second order structures.
For all $e > 1$, $\myE v_{e}(t,\theta) = \myE w_{e}(t,\theta) = 0$.

All that remains to be specified are the covariance functions on the
EOF loading processes. We use a separable Mat\'ern correlation structure
[\citet{guttorp2006}]. The Mat\'ern correlation is defined as
\[
\mathrm{M}_\nu(h/\lambda) = \frac{2^{1-\nu}}{\Gamma(\nu)}\bigl(|h/\lambda
|\bigr)\mathrm{K}_\nu\bigl(|h/\lambda|\bigr),
\]
where $h\in\real$, $\nu>0$ is the smoothness parameter and $\lambda>0$
is the range parameter.
The model correlation is
\begin{eqnarray*}
&&C(t_1,t_2,\theta_1,\theta_2;
\lambda_\alpha,\lambda_\beta,\lambda_R,
\lambda_t)
\\
&&\qquad =\M_2 \biggl( \frac{\alpha_1-\alpha_2}{\lambda_\alpha} \biggr)
\M_2 \biggl(
\frac{\beta_1-\beta_2}{\lambda_\beta} \biggr) \M_2 \biggl(
\frac{R_1-R_2}{\lambda_R} \biggr)
\M_2 \biggl( \frac{t_1-t_2}{\lambda_t} \biggr),
\end{eqnarray*}
where we fix the Mat\'ern smoothness at 2. A process with
Mat\'ern correlation with a smoothness of 2 has realizations that are almost
twice differentiable; in particular, this imposed assumption aligns with
the evolution of the geomagnetic storm across time, as a smoothly
varying process. Second, numerical
model output typically smoothly varies with input setting, and
researchers in the computer experiments literature often use a Gaussian
correlation function $C(h) = \exp(-|h|^2)$, which coincides with the
Mat\'ern class with infinite smoothness. However,
it is well known that these Gaussian correlation functions lead to
numerically poorly
behaved covariance matrices, and, in fact, researchers often add an
artificial ridge to the covariance matrix for stability.
The smoothness of a spatial process is difficult
to estimate, and using a fixed smoothness of 2 on the coefficient
processes implies model output varies smoothly between input settings.
The model is completed by specifying the covariance functions of the
EOF loadings as
%
%
\begin{equation}
\label{eqcovariance} \Cov\bigl(v_{e}(t_1,
\theta_1),v_{e}(t_2,\theta_2)
\bigr) = \sigma_{e}^2 C(t_1,t_2,
\theta_1,\theta_2; \lambda_{\alpha e},
\lambda_{\beta e}, \lambda_{R e},\lambda_{te}).
\end{equation}
The same separable covariance model is assumed for the $w_{e}$ coefficients,
but with distinct parameters. Notice that although we use a separable
structure for the coefficient processes at each level of EOF, the final
statistical model is not separable, but rather has a covariance function
that is a weighted sum of separable covariances; this class of covariances
is a type of well established product-sum covariances [\citet{decesare2001,
deiaco2001}].

%
%
\begin{table}[b]
\caption{Parameters for the mean function of $v_1(t,\theta)$ and
separable Mat\'ern covariance
functions for all EOF coefficient processes, as estimated by ordinary
least squares
and maximum likelihood, respectively. Ranges of $\alpha, \beta$, and $R$
have been standardized to $[0,1]$ for this~table}\label{tabests}
\begin{tabular*}{\textwidth}{@{\extracolsep{\fill}}lccccc@{}}
\hline
& \multicolumn{1}{c}{$\bolds{\gamma_0}$} & \multicolumn
{1}{c}{$\bolds{\gamma_1}$} &
\multicolumn{1}{c}{$\bolds{\gamma_2}$} & \multicolumn{1}{c}{$\bolds
{\gamma_3}$} & \multicolumn{1}{c@{}}{$\bolds{\gamma_4}$} \\
\hline
$\myE v_1(t,\theta)$ & 16.0 & 180 & 2804 & $-0.201$ & 16.6 \\
\hline\\
& \multicolumn{1}{c}{$\bolds{\sigma}$} & \multicolumn{1}{c}{$\bolds
{\lambda_\alpha}$} & \multicolumn{1}{c}{$\bolds{\lambda_\beta}$} &
\multicolumn{1}{c}{$\bolds{\lambda_R}$} &
\multicolumn{1}{c@{}}{$\bolds{\lambda_t}$} \\
\hline
$v_1(t,\theta)$& 11.2 & 0.22 & 0.19 & 0.1\phantom{0} & 0.051 \\
$v_2(t,\theta)$& 88.8 & 3.10 & 0.08 & 0.1\phantom{0} & 0.248 \\
$v_3(t,\theta)$& 80.1 & 2.58 & 0.24 & $10^{-3}$ & 0.200 \\[3pt]
$w_1(t,\theta)$& 24.5 & 1.05 & 0.58 & 0.01 & 0.067 \\
$w_2(t,\theta)$& 18.7 & $10^{-3}$ & 0.03 & 3.21 & 0.046 \\
$w_3(t,\theta)$& 16.9 & 0.17 & $10^{-6}$ & 5.98 & 0.035 \\
$w_4(t,\theta)$ & 15.3 & 0.18 & 1.52 & 0.02 & 0.028 \\
\hline
\end{tabular*}
\end{table}
%

\subsection{Estimation}

The main parameters of interest are the input parameters
$\theta=(\alpha,\beta,R)$, and
all other statistical parameters, such as mean function coefficients
and covariance
function ranges and variances, are of secondary interest. \citet{bayarri2007}
argue that the uncertainty in these secondary parameters is typically
substantially less than the uncertainty in the input parameters,
so that fixing the statistical parameters is justifiable in practice.
In this light, we take an empirical Bayes approach to uncertainty
quantification, where
the mean function parameters of the EOF loading processes are estimated by
ordinary least squares (OLS), and the remaining covariance function
parameters are
estimated by maximum likelihood (ML), conditional on the mean estimates.
The observational error is taken to be $5\%$ of the empirical standard
deviation of energy observations, aligning with our collaborators'
expert knowledge of the typical observational error for this type of
data set.

Table~\ref{tabests} displays the OLS estimates of the mean function
parameters and ML estimates of the separable Mat\'ern covariance function
parameters.
Recall the results of \citet{higdon2008} in that the inclusion
of higher order principal component terms typically does not assist in
prediction. As anticipated with a basis decomposition, the low order
coefficients have more variability than the high order coefficients
(noting that much of the variability of $v_1$ is accounted for in the
nonstationary mean function). The input parameters in Table~\ref{tabests}
have been standardized to the unit interval to ease comparisons between
input parameter, and we see that the greatest correlation for the
low fidelity decomposition is across the $\alpha$ index, with $\beta$
and $R$
on the same order of correlation decay. The discrepancy function,
on the other hand,
tends to be more highly controlled by the $R$ index, with $\alpha$ and
$\beta$
sharing approximately the same decay rate of correlation on average.
This indicates that, while there is some information regarding $\beta$
contained in the energy model output, there is substantially more for
$\alpha$ and $R$, which is expected, recalling the physical equation
(\ref{eqabr}).

Fixing the mean and covariance estimates, we impose independent uniform
priors on $\alpha, \beta$, and $R$, with uniformity over the bounding
boxes described at the head of this section. Define the following vectors:
\begin{eqnarray*}
\bY(t) &= &\bigl(Y(\bs_1,t),Y(\bs_2,t),\ldots,Y(
\bs_{n_o},t)\bigr)',
\\
\bH(t,\theta) &=& \bigl(H(\bs_1,t,\theta), H(\bs_2,t,
\theta), \ldots, H(\bs_{n_s},t,\theta)\bigr)',
\\
\bL(t,\theta) &=& \bigl(L(\bs_1,t,\theta),L(\bs_2,t,
\theta), \ldots, L(\bs_{n_s},t,\theta)\bigr)',
\end{eqnarray*}
where $n_o=170$ is the number of locations of observations; note we
implicitly order the observations and model output (and corresponding EOFs)
such that the first $n_o$ entries are the shared locations between the
observations and model output, and the last $n_o+1$ to $n_s$ entries
of $\bH(t,\theta)$ and $\bL(t,\theta)$ are the model output
locations with
no corresponding observations. Then combine these vectors into
\begin{eqnarray*}
\bY&=& \bigl(\bY(t_1)',\bY(t_2)',
\ldots,\bY(t_{n_t})'\bigr)',
\\
\bH(\theta) &=& \bigl(\bH(t_1,\theta)',
\bH(t_2,\theta)',\ldots,\bH(t_{n_t},
\theta)'\bigr)',
\\
\bL(\theta) &=& \bigl(\bL(t_1,\theta)',
\bL(t_2,\theta)',\ldots,\bL(t_{n_t},
\theta)'\bigr)'.
\end{eqnarray*}
Finally, combine the high and low fidelity vectors across input settings,
\begin{eqnarray*}
\bH&= &\bigl(\bH(\theta_1)',\bH(\theta_2)',
\ldots,\bH(\theta_{5})'\bigr)',
\\
\bL&=& \bigl(\bL(\theta_1)',\bL(\theta_2)',
\ldots,\bL(\theta_{20})'\bigr)'.
\end{eqnarray*}
Then $\bZ= (\bY',\bH',\bL')'$ is viewed as a realization
from the stochastic process defined by (\ref{eqobs}), (\ref{eqreslink}),
(\ref{eqLEOF}) and (\ref{eqdeltaEOF}).
Conditional on the realization $\bZ$, the posterior distribution of
$\theta$ is sampled using a Metropolis--Hastings algorithm by block
updating the vector $\theta$ at each step.
In particular, we use independent normal proposal densities centered at
the current
MCMC sample, with standard deviation one-tenth of the standard
deviation of the
initial design points (over $\theta$).

Computation of the density of $\bZ$ is difficult due to the large
dimension; for our
initial design and observations $\bZ$ is of length 748,260.
Utilizing Result 1 from \citet{higdon2008} alleviates this problem. In
particular,
\citet{higdon2008} suppose $\mathbf{x}\sim\mathrm{N}(\mathbf{0},\Sigma_x)$ and
$\bolds{\xi} \sim\mathrm{N}(\mathbf{0},\Sigma_{\xi})$ are
independent.
Let $\bZ= \bU\mathbf{x}+ \bolds{\xi}$, and
define $\hat{\bolds{\beta}}
= (\bU' \Sigma_{\xi}^{-1} \bU)^{-1} \bU' \Sigma_{\xi}^{-1} \bZ$.
Then the likelihood function of $\bZ$ can be written
%
%
\begin{eqnarray}
\label{eqresult1} L(\bZ) &\propto&| \Sigma_{\xi} |^{-1/2} \bigl|
\bU' \Sigma_{\xi}^{-1} \bU\bigr|^{-1/2}
\nonumber
\\[-8pt]
\\[-8pt]
\nonumber
&&{}\times \exp
\bigl( -\tfrac{1}{2} \bZ'\bigl(\Sigma_{\xi}^{-1}
- \Sigma_{\xi}^{-1} \bU\bigl(\bU'
\Sigma_{\xi}^{-1} \bU\bigr)^{-1} \bU'
\Sigma_{\xi}^{-1}\bigr)\bZ\bigr) L(\hat{\bolds{\beta}}).
\end{eqnarray}

In our case, $\bU$ is a block diagonal matrix of EOFs, with $1 + 5 + 20$
blocks. The very first block corresponds to the observations and is
itself a block diagonal matrix with
$n_t$ identical blocks, each of which contains the truncated EOFs
corresponding to the observation locations:
\begin{eqnarray*}
\pmatrix{ u_{L1}(
\bs_1) & \cdots& u_{L n_L}(\bs_1) &
u_{\delta1}(\bs_1) & \cdots& u_{\delta n_\delta}(
\bs_1)
\vspace*{2pt}\cr
\vdots& \vdots& \vdots& \vdots& \vdots& \vdots
\vspace*{2pt}\cr
u_{L1}(\bs_{n_o}) & \cdots& u_{L n_L}(
\bs_{n_o}) & u_{\delta1}(\bs_{n_o}) & \cdots&
u_{\delta n_\delta}(\bs_{n_o})},
\end{eqnarray*}
so that the first block of $\bU$ has dimension $(n_t \times n_o)
\times(n_t \times(n_L + n_\delta))$, in our case $(18 \times170)
\times(18 \times(3 + 4)) = 3060 \times126$. The next 5 blocks of
$\bU$
correspond to the high resolution model output, and again contain
$n_t$ blocks of EOF matrices, each of which is
\begin{eqnarray*}
\pmatrix{ u_{L1}(
\bs_1) & \cdots& u_{L n_L}(\bs_1) &
u_{\delta1}(\bs_1) & \cdots& u_{\delta n_\delta}(
\bs_1)
\vspace*{2pt}\cr
\vdots& \vdots& \vdots& \vdots& \vdots& \vdots
\vspace*{2pt}\cr
u_{L1}(\bs_{n_s}) & \cdots& u_{L n_L}(
\bs_{n_s}) & u_{\delta1}(\bs_{n_s}) & \cdots&
u_{\delta n_\delta}(\bs_{n_s})}.
\end{eqnarray*}
Hence, each of these 5 blocks of $\bU$ is of dimension $(n_t \times n_s)
\times(n_t \times(n_L + n_\delta))$, in our case $29\mbox{,}808 \times126$.
The final 20 blocks
of $\bU$ correspond to the low fidelity model output, each of which is
a block diagonal matrix consisting of $n_t$ blocks of the following EOF
matrices:
\begin{eqnarray*}
\pmatrix{ u_{L1}(\bs_1)
& \cdots& u_{L n_L}(\bs_1)
\vspace*{2pt}\cr
\vdots& \vdots& \vdots
\vspace*{2pt}\cr
u_{L1}(\bs_{n_s}) & \cdots& u_{L n_L}(
\bs_{n_s})}.
\end{eqnarray*}
Thus, each of the last 20 blocks of $\bU$ is of dimension $(n_t \times n_s)
\times(n_t \times n_L)$, in our case $29\mbox{,}808 \times54$.

The entries of $\mathbf{x}$ are EOF weights
$v_e(t,\theta)$ and $w_e(t,\theta
)$. As with
the matrix $\bU$, it is convenient to divide $\mathbf{x}$ into $1 + 5 + 20$ segments.
The first segment consists of the observation EOF coefficients
\[
\bigl(\mathbf{v}(t_1,\theta_0)',
\bw(t_1,\theta_0)',\ldots,
\mathbf{v}(t_{n_t},\theta_0)',\bw(t_{n_t},
\theta_0)'\bigr)',
\]
where
\begin{eqnarray*}
\mathbf{v}(t,\theta) &= &\bigl(v_1(t,\theta),\ldots,v_{n_L}(t,
\theta)\bigr)',
\\
\bw(t,\theta) &=& \bigl(w_1(t,\theta),\ldots,w_{n_\delta}(t,
\theta)\bigr)'.
\end{eqnarray*}
The following 5 segments correspond to the high fidelity runs, each of
which consists of
\[
\bigl(\mathbf{v}(t_1,\theta_p)',
\bw(t_1,\theta_p)',\ldots,
\mathbf{v}(t_{n_t},\theta_p)',\bw(t_{n_t},
\theta_p)'\bigr)'
\]
for $p=1,\ldots,5$. The final 20 segments correspond to the low fidelity
runs and consist of
\[
\bigl(\mathbf{v}(t_1,\theta_p)',\ldots,
\mathbf{v}(t_{n_t},\theta_p)'\bigr)'
\]
for $p=1,\ldots,20$. Note that Result 1 of \citet{higdon2008} requires
$\mathbf{x}$ be centered at zero; to this end, we
apply Result 1 to
$\bZ-\bU\myE\mathbf{x}= \bU(\mathbf{x}-\myE\mathbf{x}) + \bolds{\xi}$.

Similar to $\bU$ and $\mathbf{x}$, we break up $1
+ 5 + 20$ segments of
$\bolds{\xi}$. The first $n_t \times n_o$
have variances $\tau^2 + \tau_L^2 + \tau_\delta^2$;
the following $5 \times n_t \times n_s$ have variances $\tau_L^2 +
\tau
_\delta^2$
and the remaining $20 \times n_t \times n_s$
entries have variances $\tau_L^2$. This completes our model's
formulation of the
likelihood decomposition of Result 1 of \citet{higdon2008}.

Exploiting the EOF decomposition of the model output dramatically reduces
dimensionality of the problem. For example, a typical Gaussian process
approach to
our setup would require inverting a matrix of dimension $748\mbox{,}260 \times
748\mbox{,}260$, whereas,
for example, inverting $\bU' \Sigma_{\varepsilon} \bU$ is feasible,
as it
is a matrix of dimension $1836 \times1836$.

\section{Results and sequential design} \label{secresultsEI}

\subsection{Initial calibration}

\begin{figure}

\includegraphics{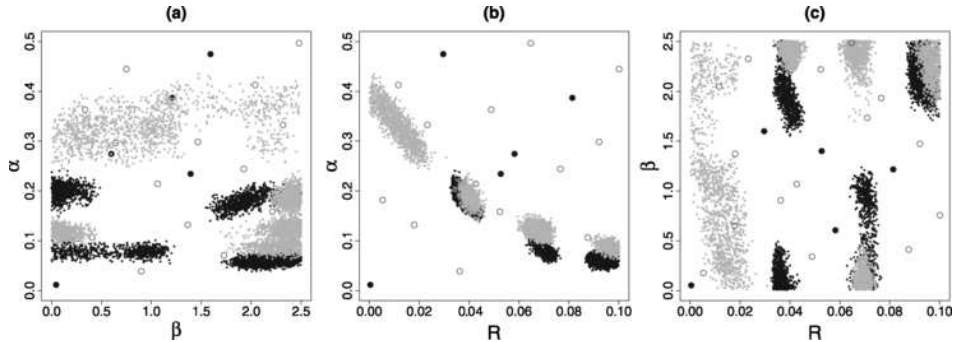}

\caption{Posterior samples using only the five high fidelity runs
(small grey dots), and using the entire initial design of five high fidelity
and 20 low fidelity runs (small black dots) with
input pairs at which the low fidelity model was run (unfilled circles)
and input pairs at which both low and high fidelity models were run
(filled circles).}\label{figinitial}
\end{figure}

Initially, we begin by running five independent chains of posterior samples
simultaneously, from random starting values. The posterior samples
based on
the initial design are shown
in Figure~\ref{figinitial} as small black dots.
Notice the distribution is multimodal, and there is an apparent nonlinear
inverse relationship between $\alpha$ and $R$. In fact, the curve along
which the posterior
samples fall for $(\alpha, R)$ define a posterior distribution of
total energy.
Recall equation (\ref{eqmeanfunction}), where we exploited the functional
form of total energy, of a form $\alpha+ R\sqrt{\alpha}$. These results
suggest that the quantity of total energy is well defined based on our
observations and initial design, and a combination of pairs of input
parameters $(\alpha,R)$ that approximately yield this total energy are
appropriate for our data set. Notice that $\beta$
is not especially well identified based on our observations. This is expected,
as we currently are modeling only energy, and $\beta$ is a controlling
parameter
for flux, although the information in the energy variable regarding
$\beta$ is not negligible.

%
%


Let us illustrate the benefit of using the low fidelity model in conjunction
with the high fidelity model. If there were no extra information added by
including the low fidelity model output, we would expect the posterior
samples based exclusively on the high fidelity version to be the same
as including both model fidelities. The small grey dots of Figure~\ref{figinitial} are posterior samples for the input
parameters based on only the five high fidelity runs, here ignoring the
20 low fidelity runs. In particular, the statistical model remains the same,
except where we write
%
%
\begin{equation}
\label{eqHEOF} H(\bs,t,\theta) = \sum_{e=1}^{n_H}
u_{He}(\bs) v_{e}(t,\theta) + \varepsilon_{H}(
\bs,t,\theta),
\end{equation}
where $n_H = 3$, and $\myE v_1(t,\theta)$ has the same functional form
as (\ref{eqmeanfunction}).
Comparing the two sets of posterior samples in Figure~\ref{figinitial}
shows the gain in augmenting the high fidelity runs with the low fidelity
information---the location of the curve in panel~(b) for the pair
$(\alpha,R)$
is adjusted downward
when also using the low fidelity runs and a posterior mode is ruled out.
Specifically, the posterior mode about $(\alpha,R) \approx(0.35,0.01)$
is no longer present. Hence, our posterior uncertainty regarding the
parameters $\alpha$ and $R$ has decreased due to the inclusion of the
low fidelity output. The posterior samples for $\beta$ are slightly
adjusted when the low fidelity information is included,
although not necessarily the same amount as for $\alpha$ and $R$,
again, due to the fact that $\beta$ is linked to flux.

There are two potential explanations for the multimodal
nonlinear behavior of the posterior
distribution shown in Figure~\ref{figinitial}(b).
The first is that the observations have no
information regarding the specific pair of $(\alpha, R)$ that is optimal
or, alternatively, the curve is an artefact of the sparse
initial design. In particular, with only 5 runs of the high fidelity model,
it is unlikely that the discrepancy function $\delta(\bs,t,\theta)$
has been
well estimated, and given more runs of the LFM--MIX, the posterior distribution
may shrink to one of the modes of Figure~\ref{figinitial}. To this end,
we develop a
sequential design based on expected improvement.

\subsection{Expected improvement for sequential design}

We seek to perform an additional run of the LFM--MIX based on current
information,
and expected improvement (EI) is one approach to sequential design that
incorporates
accuracy and uncertainty. Expected improvement was originally developed for
black-box function optimization [\citet{jones1998}], but we adjust the
idea for our purposes of
parameter identification. To begin, we define the improvement function
for a
given location and time as minimizing the squared residual between the high
fidelity model output and observations:
%
%
\begin{equation}
\label{eqI} I(\bs,t,\theta) = \max\bigl\{f_\mathrm{min} - \bigl
(Y(\bs,t) - H(
\bs,t,\theta)\bigr)^2, 0 \bigr\},
\end{equation}
where $f_\mathrm{min} = \min_{i=1}^{5} (Y(\bs,t) - H(\bs,t,\theta_i))^2$
is the observed
minimized squared residual over the initial runs of the LFM--MIX.
The EI is defined as a sum of expected improvement
functions over all locations and times,
%
%
\begin{equation}
\label{eqEIsum} \mathit{EI}(\theta) = \sum_{\mathbf{s},t} \myE I(
\bs,t,\theta),
\end{equation}
and is a function only of input parameter $\theta$.

To write the closed form of EI at an arbitrary setting
$\theta$, we require the conditional distribution of the high fidelity
model, given the current runs. In particular, we have
%
%
\begin{equation}
H(\bs,t,\theta) | \bigl\{ H(\bs,t,\theta_i)\bigr
\}_{i=1}^5, \bigl\{L(\bs,t,\theta_i) \bigr
\}_{i=1}^{20} \sim\mathrm{N}\bigl(\hat{H},\hat{
\sigma}^2\bigr),
\end{equation}
where $\hat{H}$ and $\hat{\sigma}^2$ are simply a conditional mean and
variance of
the multivariate normal defined by equations (\ref{eqreslink}), (\ref
{eqLEOF}),
and (\ref{eqdeltaEOF}).
Let $Q_{\pm} = (Y - \hat{H} \pm\sqrt{f_\mathrm{min}})/\hat{\sigma}$,
we simplify notation by setting $Y = Y(\bs,t)$ and $\phi$ and $\Phi$
are the standard
normal density and cumulative distribution functions, respectively.
Then the expected improvement at location $\bs$ and time $t$ has
closed form
%
%
\begin{eqnarray}
\label{eqEIclosed}\qquad \myE I(\bs,t,\theta) &=& \bigl(f_\mathrm{min} - (Y -
\hat{H})^2 - \hat{\sigma}^2\bigr) \bigl(\Phi(Q_+) -
\Phi(Q_-)\bigr)
\nonumber
\\[-8pt]
\\[-8pt]
\nonumber
&&{}+ \hat{\sigma} \bigl( (\sqrt{f_\mathrm{min}} + \hat{H} - Y)\phi
(Q_+) + (
\sqrt{f_\mathrm{min}} + Y - \hat{H})\phi(Q_-) \bigr).
\end{eqnarray}
See the \hyperref[app]{Appendix} for a derivation. Notice that
EI is indeed a weighting between uncertainty $(\hat{\sigma})$
and accuracy $((Y-\hat{H})^2)$. For example, if, at a new setting
$\theta$,
our predictive variance for the high fidelity model output was small,
then the latter term of (\ref{eqEIclosed}) will be negligible, and the
EI will be controlled by the accuracy in the first term
as a function of $(Y - \hat{H})^2$.

%
%
\begin{figure}

\includegraphics{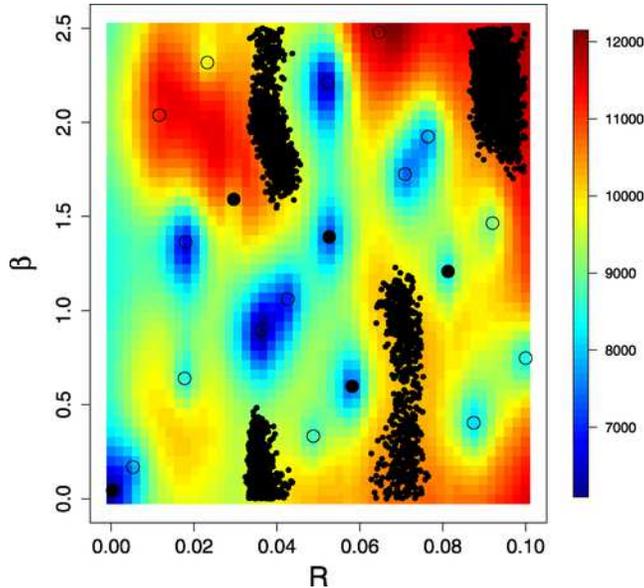}

\caption{Expected improvement surface, with initial posterior samples
(small dots) based on initial design over $\theta$ with
input pairs at which the low fidelity model was run (unfilled circles)
and input pairs at which both low and high fidelity models were run
(filled circles).}\label{figEI}
\end{figure}


Figure~\ref{figEI} shows the EI surface as a function of $\beta$
and $R$ for the best value of $\alpha$ ($0.5$). As previously, the
open circles
are locations at which we ran the low fidelity model, and the closed circles
are the locations at which we ran both fidelities. There
are a number of interesting features illustrated by this surface.
The EI surface is multimodal, with the most pronounced mode at $(\beta
,R) =
(2.5,0.068)$, falling directly between two modes of the initial
posterior samples. In this area, the uncertainty is substantial
enough that an optimum may be in the area. Note
there are no high fidelity model runs in the immediate area; that
the EI maximum also falls directly between two posterior sample modes
indicates that EI is indeed a weighting between uncertainty and
accuracy. EI is sensitive to the initial design, and at most of the
locations where the low or high fidelity model was run, there are
relatively low values of EI, as we have already reduced our uncertainty
in those areas. However, the EI surface also follows the general
trend of the initial posterior samples, indicating our initial samples
fell in areas of high model accuracy.

\begin{figure}

\includegraphics{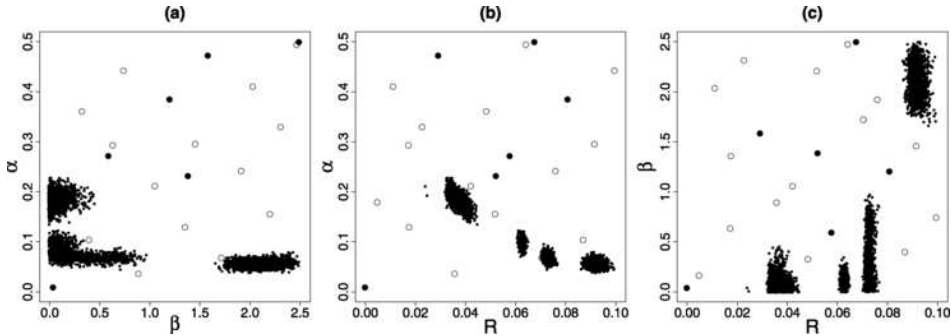}

\caption{Second round of posterior samples (small dots) based on
initial design plus the run suggested by the expected improvement
criterion with
input pairs at which the low fidelity model was run (unfilled circles)
and input pairs at which both low and high fidelity models were run
(filled circles).}\label{figsecond}
\end{figure}

We ran the high and low fidelity version of the LFM--MIX at the greatest
mode indicated by the EI surface, specifically at
$(\alpha,\beta,R) = (0.5, 2.5, 0.068)$, and conditional on this additional
run, sampled from the posterior distribution of the input parameters.
If no extra information were added due to the sequential
design run, we would see the same posterior samples as in Figure~\ref{figinitial}.
The second round of posterior samples, conditional on the initial
design plus the
single additional run suggested by EI, are shown in Figure~\ref{figsecond}. The substantial change between Figures~\ref{figinitial}
and \ref{figsecond} can be seen in the third panel (c), the pairwise posterior
samples for $\beta$ and $R$. In particular, the upper leftmost mode
that was present
in Figure~\ref{figinitial}(c) has been ruled out now, as there are
no posterior samples in this area. Our posterior uncertainty has decreased
due to the single additional run suggested by EI.
Our information regarding $R$ has also increased due to the added
EI run, as the initial middle mode about $R=0.7$ has now split into
two smaller modes.

In previous experiments with the LFM--MIX,
continuing sequential design based on EI
improves the posterior distribution of $(\alpha,R)$ slowly
and primarily explores the three-dimensional $(\alpha,\beta,R)$ space
over $\beta$.
This reiterates the substantial uncertainty in $\beta$ based on the energy
variable alone, and, unfortunately, due to the high budgetary demand
of running the LFM--MIX,
at 100 hours for each run of the high and low fidelity model on 8 processors,
it is not within our current budget to continue the sequential design.
Future work is aimed at including observations for flux, which we anticipate
greatly improving identification of $\beta$.

%
%


\section{Parameter estimation for the Lorenz `96} \label{seclorenz}

In the previous section we outlined a statistical model for combining high
and low fidelity model output for large spatiotemporal data sets with an
application of quantifying the uncertainty in input parameters for
the LFM--MIX computer model. The initial posterior
distributions illustrated a strong nonlinear relationship between the parameters
$\alpha$ and $R$, and based on a sequential design framework, we saw
the posterior distributions shrink in variability, ruling out an
area of the parameter space present in the
initial multimodal posterior distribution.
In this section we illustrate a similar statistical model
using a physical model with known truth. The goal in this section
is to compare our ability to identify model parameters using
the EOF approximation model with differing initial design sizes, and
to assess the ability of sequential design under expected improvement
in improving the posterior estimates of unknown parameters.

The Lorenz `96 system (hereafter L96) of equations was developed by
Edward Lorenz to be a simplified
one-dimensional atmospheric model that exhibits chaos [\citet
{lorenz1996}]. The
physical model is for 40 variables (known as state variables in the atmospheric
sciences). For variable $Y(\bs,t)$, location $\bs= 1,\ldots,40$ and
time~$t$, we have
%
%
\begin{eqnarray}
\mathrm{d}Y(\bs,t)/ \mathrm{d}t& =& -Y(\bs-2,t)Y(\bs-1,t) + Y(\bs-1,t)Y(\bs
+1,t)
\nonumber
\\[-8pt]
\\[-8pt]
\nonumber
&&{} - Y(
\bs,t) + F(\bs),
\end{eqnarray}
where $F(\bs)$ is a location dependent forcing term, and $Y(\bs,t)$
is available
at any integer value of $\bs$ by setting $Y(\bs-40,t) = Y(\bs+40,t) =
Y(\bs,t)$.
For the forcing term, \citet{lorenz1996} used $F(\bs) = 8$,
but for our purposes we wish to mimic the behavior of the LFM--MIX using
this reduced atmospheric model.

Analogous to the LFM--MIX case, we have two forcing functions, corresponding
to a low and a high fidelity simulator. In particular, we,
respectively, define the
low and high fidelity forcing functions as
%
%
\begin{eqnarray}
\label{eqfl} F_L(\bs;a,b) &=& 8 + a + 3ab \exp\bigl(-\cos(2\pi
\bs/40)\bigr)/\exp(1),
\\
\label{eqfh} F_H(\bs;a,b) &=& 8 + a + 3ab \exp\bigl(-10\cos(2\pi
\bs/40)\bigr)/\exp(10).
\end{eqnarray}
Notice the functional form here, $a + ab$, is akin to the total energy
equation of the LFM--MIX, which was of the form $\alpha+ R\sqrt
{\alpha}$.

\begin{figure}

\includegraphics{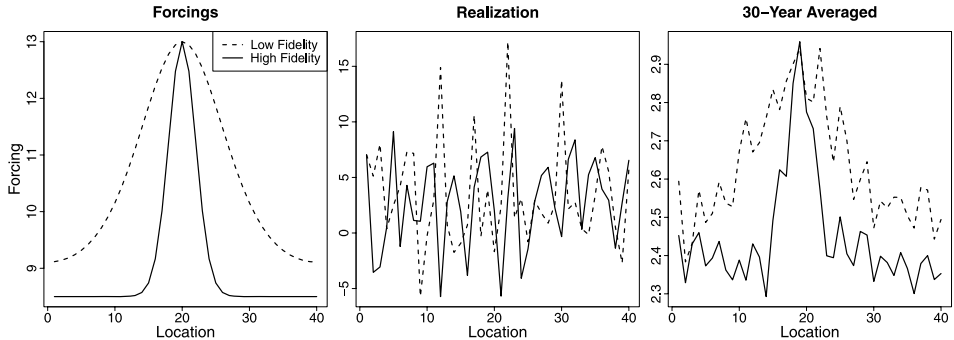}

\caption{Illustration of the Lorenz `96 model. Forcings for the low
and high
fidelity versions, physical model realizations, and a 30-year averaged
run. Forcings correspond to $a=1/2$ and $b=3$.}\label{figlorenz}
\end{figure}

Fixing true values of $a$ and $b$ at $1/2$ and $3$, respectively,
the first panel of Figure~\ref{figlorenz} shows the corresponding forcing
functions for the low and high fidelity versions. Notice the low fidelity
version appears to smear out the peak defined by the high fidelity
forcing function.
This is akin to the relationship between the differing fidelities of
the LFM--MIX,
where the low fidelity model tends to produce output that
is a (spatially) less peaked version of the more peaked high fidelity model
output.

%
%


The observations are generated from the high fidelity version of the L96,
based on 40 independent initial unit uniform random variables. Solving
the equations
every 6 hours, we run the L96 for 300 years, and use 30-year averaged
output, garnering approximate climate of the L96.
The motivation for time-averaging is that each single realization from
the L96
is highly erratic, as seen in Figure~\ref{figlorenz}, whereas taking
time-averages over long periods tends to reproduce the forcing
function, also displayed in Figure~\ref{figlorenz}. To these 10
time realizations, we add independent normal errors for each variable
at all time points, whose mean is zero and whose standard deviation is
five percent of the empirical standard deviation of the model output,
again to line up with the expert understanding of measurement error
for the LFM--MIX example.

We suppose it is known that $a \in[0,2]$ and $b \in[0,5]$. To explore
different design approaches, we run two initial designs. The first design
assumes greater resources than are available for the LFM--MIX.
In this situation, we run the low fidelity model at 40 pairs of
input settings based on a space-filling design, and the high fidelity
model at a
space-filled subset at 20 points of the original 40. This setup is
designed is to illustrate our ability to tune model parameters
in the situation with more resources than are currently available.
The second design utilizes a space-filled subset of 20 runs of the low
fidelity computer model, with an additional 5 runs of the high fidelity
version, aligning directly with our setup for the LFM--MIX scenario.

To align with the LFM--MIX modeling approach, we suppose the
observations are
adequately represented by the high fidelity version of L96, up to white
noise. In particular, using similar notation as
in the previous section where $\theta= (a,b)$, we write
%
%
\begin{equation}
\label{eqobslorenz} Y(\bs,t) = H(\bs,t,\theta_0) + \varepsilon(
\bs,t),
\end{equation}
where $\varepsilon(\bs,t)$ is a white noise process, which we assume
to be
normally distributed with mean zero and variance $\tau^2$. As with the
LFM--MIX,
we link the low and high fidelity models with an additive discrepancy function
$\delta(\bs,t,\theta)$, where
%
%
\begin{equation}
\label{eqreslinklorenz} H(\bs,t,\theta) = L(\bs,t,\theta) + \delta
(\bs,t,
\theta).
\end{equation}

Whereas the LFM--MIX is highly multivariate, our L96 example does not
require the same dimension reduction techniques employed earlier.
Although not required, we use similar modeling techniques to those employed
for the LFM--MIX above in order to explore our ability to
identify\vadjust{\goodbreak}
physical parameters in a setting where approximations are required.
Hence, we write
\[
L(\bs,t,\theta) = \sum_{e=1}^{n_L}
u_{Le}(\bs) v_{e}(\theta,t) + \varepsilon_{L}(
\bs,t,\theta)
\]
and
\[
\delta(\bs,t,\theta) = \sum_{e=1}^{n_\delta}
u_{\delta e}(\bs) w_{e}(\theta,t) + \varepsilon_{\delta}(
\bs,t,\theta).
\]
Putting $n_L = 2$ and $n_\delta= 1$ (capturing more than $99\%$ of
the variability), the residual processes
$\varepsilon_L$ and $\varepsilon_\delta$ are modeled as normally
distributed white noise terms with variances $\tau_L^2$ and $\tau
_\delta^2$,
respectively. As in the LFM--MIX case, we model $v_{1},v_{2}$, and
$w_{1}$ as Gaussian processes. Each is endowed with a mean
function of the form $\gamma_0 + \gamma_1 a + \gamma_2 b\sqrt{a}$,
a functional form that was decided upon after elementary data analysis;
notice we find similar behavior to the $a + ab$ form of the
forcing functions (\ref{eqfl}) and (\ref{eqfh}).
Unlike the LFM--MIX, we suppose the $v$ and $w$ processes are independent
across time; indeed, with the L96, we consider long term averages, and
viewing the realizations as independent across time is justifiable,
whereas in the LFM--MIX case, our realizations arise from a continuous
process over a relatively short time interval. The functional form
of the covariance for the $v$ and $w$ coefficient processes is
$\sigma^2 C(\theta_1,\theta_2;\lambda_a,\lambda_b)$, where
$\theta=(a,b)$, and
\[
C(\theta_1,\theta_2;\lambda_a,
\lambda_b) = \M_2 \biggl( \frac{a_1-a_2}{\lambda_a} \biggr)
\M_2 \biggl( \frac{b_1-b_2}{\lambda_b} \biggr),
\]
where naturally each $v_1, v_2$, and $w_1$ has distinct covariance
and regression parameters.

For physical parameter estimation, we sample the posterior distribution of
$\theta$ conditional on $\bZ$, which is made up of the following
components. Define the vectors
$\bY(t_i) = (Y(\bs_1,t_i),Y(\bs_2,t_i),\ldots,Y(\bs_{n_s},t_i))'$,
$\bH(t_i) = (H(\bs_1,t_i,\theta_1),\break  H(\bs_2,t_i,\theta_1),
\ldots,H(\bs_{n_s},t_i,\theta_{n_H}))'$,
and $\bL(t_i) = (L(\bs_1,t_i,\theta_1),
L(\bs_2,t_i,\theta_1),\ldots,\break   L(\bs_{n_s},t_i,\theta_{n_L}))'$,
where the number of low and high fidelity samples are $n_L$ and $n_H$,
respectively. Combine these vectors into the single time point vector
$\bZ(t_i) = (\bY(t_i)',\bH(t_i)',\bL(t_i)')'$,
then $\bZ= (\bZ(t_1)',\ldots,\bZ(t_{n_t})')'$.

\begin{figure}

\includegraphics{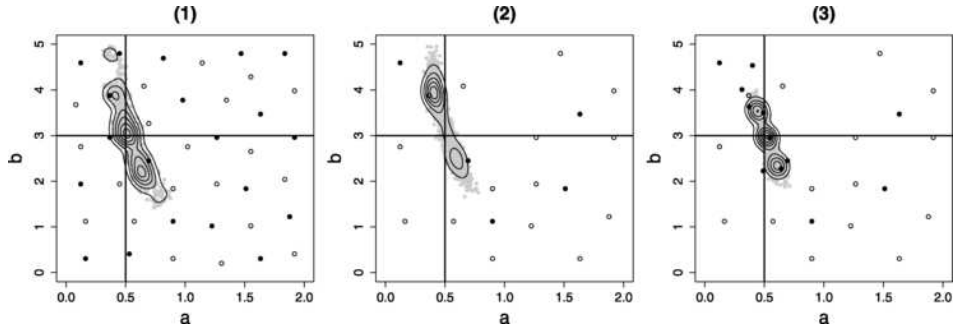}

\caption{Parameter turning the Lorenz `96 model. True parameter
values are $(a,b) = (1/2,3)$, indicated by the intersection of two
solid lines. Each panel contains posterior densities with
contours overlying posterior samples for
(1): large initial design, (2): sparse initial design similar to the
LFM--MIX, (3): sparse initial design with seven additional runs
chosen sequentially by expected improvement.
Input settings at which the low
fidelity model was run are displayed as circles both filled
and unfilled, and settings where the high fidelity model was also run
are shown as filled circles.}\label{figlorenzcalibrate}
\end{figure}

Posterior distributions are shown in Figure~\ref{figlorenzcalibrate},
with the truth indicated by the intersection of solid lines.
We consider three cases for posterior sampling---the first is based on
a dense design of $n_L = 40$ and $n_H = 20$, shown in panel (1).
The posterior distribution
covers the truth, but is spread over a swath of plausible values, falling
along a curve of the form $a + b\sqrt{a}$, exhibiting similar behavior
as the LFM--MIX; note the substantially larger initial design size, however.
The posterior mode is at approximately $(a,b) = (0.51, 3.09)$,
indicating accurate point estimation, but still displaying
substantial uncertainty.

%
%


The middle panel of Figure~\ref{figlorenzcalibrate} replicates the
situation of the LFM--MIX more closely in that we use only $n_L=20$
and $n_H = 5$ points in the initial design. The posterior distribution
covers the true value of $(a,b)$, and again we see a swath of density
following a curve similar to $a + b\sqrt{a}$. Here, however, the posterior
mode is at $(a,b) = (0.40, 3.94)$, so while the truth is indeed captured
within the posterior samples, there appears to be some bias.
Following this sparse initial sample,
we run both low and high fidelity versions of the L96 at seven additional
input settings chosen sequentially based on the expected improvement
criterion. The final panel of Figure~\ref{figlorenzcalibrate} displays
the posterior distributions based on these $n_L = 27$ and $n_H = 12$
samples. Indeed, the posterior variability has decreased as compared to
that based on the initial design, but also notice that the posterior
has substantially less variability than the dense initial sample of
panel (1). These results suggest
we can perform fewer runs initially, and rely on a sequential design
such as expected improvement to home in on the true values. The
posterior mode after sequential design is approximately
$(a,b) = (0.52, 2.96)$, indicating accurate posterior estimation.
An interesting note is that the final posterior distribution displays
three distinct modes (although the mode about the truth is of higher
posterior density). Given that the sequential design runs cover the
posterior modes, we do not anticipate the posterior distribution
improving greatly, but reiterate that the posterior distribution contains
and is indeed centered about the truth.

\section{Discussion} \label{secdiscussion}

We have introduced an approach to quantify the uncertainty about input
parameters for large spatiotemporal
data sets with high and low fidelity model outputs.\vadjust{\goodbreak} We suppose the high fidelity
model is an adequate representation of reality at some unknown set of input
parameters up to white noise.
The high and low fidelity models are linked through an additive discrepancy
function. This link allows us to run the higher cost high fidelity
model at fewer
sets of input parameters, and explore the input setting space with the cheaper
low fidelity model. In our first example we examined the LFM--MIX model for
geomagnetic storms occurring in Earth's near space environment, which
is partially
parameterized by three unknown input parameters controlling energy and
flux. Based on an initial experimental design, using observations of energy,
we discovered a nonlinear relationship between a subset of the input settings,
which was a level curve for the total energy quantity. One input setting
was not well identified, but considering that particular variable contributes
mainly to flux, it is unsurprising that it is not well identified using only
energy observations.

To improve posterior estimation, we developed an expected improvement
criterion for sequential design. The improvement function seeks to
minimize squared distance between the high fidelity
model and observations. We derived the closed form for EI
over arbitrarily many locations and times, which simultaneously weights
uncertainty and accuracy. Based on the EI criterion,
we performed an additional run of the LFM--MIX and found that the
posterior distributions for the input parameters indeed
shrunk in width. This suggests that the nonlinear behavior of the
initial posterior distribution is potentially an artefact of our
sparse initial design. Comparing these results to the contrived
Lorenz `96 system with known truth, we would anticipate some improvement
manifesting as smaller posterior variability if we were to continue
sequential design based on EI, with the posterior mode eventually
settling around the true unknown parameter value.

In a previous set of experiments, we explored sequential design based
on EI, and found that the criterion
primarily becomes overwhelmed by the uncertainty surrounding the
input parameter involving flux. Due to the high budgetary demand of
running the LFM--MIX, it is not within our current capacity to continue the
sequential design. Our current research is aimed at including
observations for flux, which we anticipate greatly improving the posterior
distributions of all three input parameters.

We reduced dimensionality of the large data set by projecting spatial fields
onto empirical orthogonal functions; the motivation was driven by
exploratory analysis where the first main mode of spatial variation
exhibited a magnitude with functional form similar to physical
equations governing energy and flux for the LFM--MIX.
In other contexts for other space--time computer models, a different
approach may be required. For instance, if the model output is a highly
nonlinear response of input parameters, a principal component approach is
likely to be unsuccessful in statistically modeling physical model output.
In such cases the practitioner may need to perform statistical tests for
space--time separability, such as those developed by
\citet{fuentes2006} or \citet{mitchell2005}.

The clearest route of future research is to develop a bivariate model for
energy and flux, which will allow us to simultaneously identify the three
parameters controlling these two distinct variables. One potential
solution to this added complication is to use a similar EOF decomposition
for flux, and use a multivariate Gaussian process representation for the
EOF coefficient processes for both energy and flux, thereby accounting
for correlation between the two distinct variables.

The statistical model did not account for systematic model bias. Our
approach is consistent with the mathematical formulation of solving
large scale inverse problems using computer models and observed data
[see, e.g., the cosmic microwave background application in \citet{higdon2011}].
With only one observed geomagnetic storm, model bias is confounded
with the residual process; with multiple storms we could potentially
include a full bias term across space and time. However, it is believed
by space physicists
that the infinite resolution version of the LFM--MIX is unbiased, and
our high fidelity version is an approximation to this infinite
resolution. The
discrepancy function we introduced connected the low and high fidelity
versions of the model, which is notably different than the
original suggestion of \citet{kennedy2001} of including an additive
model discrepancy term. In our situation, we have only
one realization of the spatiotemporal process and, hence, model bias is
unidentifiable without some simplifying assumptions (such as constancy
across time or space).
\citet{heaton2013} also examine the LFM--MIX,
taking a predictive process approach to dimension reduction [\citet
{banerjee2008}],
and assume a rotational bias across time. That is, the authors assume
there is an unknown spatial rotation at each time point that defines model
bias for the high fidelity version. Their posterior distributions
differ from those found herein, generally centering on approximately
$(\alpha,\beta,R) = (0.47, 1.59, 0.02)$. This is not
contradictory to our results in that the assumptions regarding model
bias are different---indeed, optimal parameter values under \emph{rotated}
model output are expected to be different than those under no such
rotations. With additional geomagnetic storms, our goal is to determine
the need for such rotations and potentially fully general space--time
model biases, but it is currently unclear which of these competing
assumptions is necessary.

The low and high fidelity versions of the LFM--MIX are generated by
differing resolutions of the LFM model. While in the current work we used
only two resolutions, there is potential for a higher resolution
available that is extremely computationally intense, and must be run on
a supercomputer on at least 32 processors. Potentially, one way to
include this ``highest'' fidelity
is to maintain our model's formulation, and write the high fidelity
model as a sum of the highest fidelity and a secondary discrepancy function.
It is likely that the discrepancy connecting the lower fidelities
will be correlated with the discrepancy\vadjust{\goodbreak} connecting the higher fidelities
and, hence, we anticipate requiring a multivariate Gaussian process model
for the discrepancy processes.


%
\begin{appendix}\label{app}
\section*{Appendix}

In this appendix we derive the closed form for the expected improvement at
a single location $\bs$ and time $t$, equation (\ref{eqEIclosed}).
For notational simplicity, write $Y(\bs,t) = Y$,
$H(\bs,t,\theta) = H$, and $f_{\mathrm{min}} = f$. Then we have
\begin{eqnarray*}
\myE I(\bs,t,\theta) &= &\myE\max\bigl\{f - (Y - H)^2, 0 \bigr\}
\\
&=& \int_{f > (Y-H)^2} \bigl( f - (Y-H)^2 \bigr) L(H)\, \mathrm{d}H
\\
&= &\frac{1}{\hat{\sigma}} \int_{f > (Y-H)^2} \bigl( f -
(Y-H)^2 \bigr) \phi\biggl( \frac{H-\hat{H}}{\hat{\sigma}} \biggr
)\, \mathrm{d}H
\\
&=& \int_{{(Y-\sqrt{f}-\hat{H})}/{\hat{\sigma}} < x <
{(Y+\sqrt
{f}-\hat{H})}/{\hat{\sigma}}} \bigl( f - (Y-\hat{H}-\hat{
\sigma}x)^2 \bigr) \phi(x)\, \mathrm{d}x
\\
&=& \int_{Q_-}^{Q_+} \bigl( f-(Y-
\hat{H})^2 \bigr)\phi(x) \,\mathrm{d}x + 2\hat{\sigma} (Y-\hat{H})
\int
_{Q_-}^{Q_+} x\phi(x)\, \mathrm{d}x
\\
& &{}- \hat{\sigma}^2 \int_{Q_-}^{Q_+}
x^2 \phi(x) \,\mathrm{d}x
\\
&= &A + B + C,
\end{eqnarray*}
utilizing the change of variables $x = (H-\hat{H})/\hat{\sigma}$.
The three
integrals of $A, B$, and $C$ can be written
\begin{eqnarray*}
A &=& \bigl( f-(Y-\hat{H})^2 \bigr) \bigl( \Phi(Q_+) - \Phi(Q_-)
\bigr),
\\
B &=& 2\hat{\sigma}(Y-\hat{H}) \bigl( \phi(Q_-) - \phi(Q_+) \bigr),
\\
C &=& -\hat{\sigma}^2 \bigl( Q_- \phi(Q_-) - Q_+ \phi(Q_+) +
\Phi(Q_+) - \Phi(Q_-) \bigr),
\end{eqnarray*}
using integration by parts and the fact that the antiderivative of
$x\phi(x)$ is
$-\phi(x)$. Combining terms yields (\ref{eqEIclosed}).
\end{appendix}

\section*{Acknowledgments}
We gratefully acknowledge Doug Nychka for numerous discussions and
providing the Lorenz `96 code. The National Center for Atmospheric
Research is managed by the University
Corporation for Atmospheric Research under the sponsorship of NSF.

%



\printaddresses

\end{document}